\documentclass[]{iopart}
\begin{document}

\title[]{Boundary conditions for hyperbolic formulations of the 
Einstein equations}

\author{Simonetta Frittelli\dag\
 and Roberto G\'{o}mez\ddag  
}

\address{\dag\ Department of Physics, Duquesne University,
       Pittsburgh, PA 15282}

\address{\ddag\ Pittsburgh Supercomputing center, 4400 Fifth Avenue,
	 Pittsburgh, PA 15213}

\ead{simo@mayu.physics.duq.edu}
\begin{abstract}
In regards to the initial-boundary value problem of the Einstein
equations, we argue that the projection of the Einstein equations
along the normal to the boundary yields necessary and appropriate
boundary conditions for a wide class of equivalent formulations.
We explicitly show that this is so for the Einstein-Christoffel
formulation of the Einstein equations in the case of spherical
symmetry. 

\end{abstract}


\submitto{\CQG}

\maketitle

\section{Introduction \label{sec:1}}

Achieving long time simulations of binary black hole spacetimes
remains one of the most pressing issues in numerical relativity. 
Numerical simulations are plagued with instabilities, the origin of
which has not been isolated among a number of possible factors of
relevance to stability (see for instance
~\cite{scheel,hisaaki,lehner} and references therein). A partial list
of suspected sources of instabilities includes ill-posed evolution,
ill-posed constraint propagation and poor choices of boundary
conditions, as well as poor choices of binary black hole data. The
problem of well posedness of the evolution equations and constraint
propagation in the analytic (as opposed to numerical) sense has been
widely studied (see the Living Review article by
Reula~\cite{lecolivrev}), with the result that there is a great deal
of choice of strongly hyperbolic formulations of the Einstein
equations available for analytic studies (aside from numerical
implementation). On the other hand, appropriate boundary values for
evolution remain to be identified. Since a great deal more about the
boundary value problem of a strongly hyperbolic set of equations is
known than for any other kind, currently analytic investigations of
appropriate boundary conditions tend to use some kind of strongly
hyperbolic formulation of the Einstein equations. Relevant
considerations usually run along the following lines. 

The boundary values of a strongly hyperbolic system of partial
differential equations split always into two
sets~\cite{kreissbook}: those that are determined by the initial
conditions of the problem and cannot be chosen freely, and those
that are entirely arbitrary because they are determined by values
of the fields outside of the region of interest. In practice, the
free boundary data must be specified in order for a unique
solution of the equations to be determined in the region of
interest.

Suppose we have a boundary delimiting a region where we seek a
solution to the Einstein equations in some strongly hyperbolic
formulation. In this case one calculates data on the initial slice
satisfying the constraints, and then integrates up the hyperbolic
equations step by step. But, by the previous paragraph, the
initial data alone are not sufficient to find a unique solution in
the future of a bounded region in space. Boundary values must be
prescribed as well. Suppose that we know exactly which variables'
boundary values are to be considered independent of the initial
data. How are we to prescribe them? Since in some sense the
solution that we are seeking must satisfy all the Einstein
equations, and by construction the evolution equations
are being imposed, then it seems that the appropriate question to
ask is: What are the boundary data that preserve the constraints? 

This question is addressed by Stewart in \cite{stewart98} within a
particular strongly hyperbolic formulation of the Einstein
equations~\cite{simo94,simo96}. The constraints themselves,
considered as real functions of spacetime, propagate according to a
strongly hyperbolic system of their own. This implies that vanishing
values of the constraints at the initial time will propagate towards
the future along characteristics.  Some of the  constraints
propagate towards the boundary and cross out of the region of
interest, whereas others propagate into the region of interest by
crossing the boundary from outside. Clearly the values of the
incoming constraints at the boundary are arbitrary, and one wants to
have them vanish. But the vanishing of the constraints cannot be
imposed along the boundaries in practice. The constraints involve
derivatives of the fields across the boundaries, not just the values
of the fields themselves. Stewart argues that the vanishing of the
incoming constraints can in fact be used as a boundary condition on
the fields when the equations are linearized around flat space. If
the fields are expressed in integral form in terms of
Fourier-Laplace transforms, the linearity of the differential
equations implies algebraic equations for the transforms of the
fields.  Additionally, the constraints transform into algebraic
expressions in terms of the transforms of the fields, thus acquiring
an algebraic look. Regardless of whether these seemingly algebraic
conditions are practical boundary conditions for a numerical
simulation, the argument does not hold up in the non-linear case. 

The idea of imposing the vanishing of the incoming constraints as
boundary conditions is pursued further in \cite{calabrese01}, where
space derivatives of the fields are eliminated in favor of time
derivatives in the expression of the incoming constraints in terms
of the fundamental variables. In this case, a different formulation
of the Einstein equations is used \cite{fixing}, restricted to
spherical symmetry.

Clearly not all formulations have the problem of imposing the
constraints at the boundary. Suppose there is a formulation of
the Einstein equations that has no incoming constraints, that is:
a formulation in which the constraints propagate upwards along
the boundary. In such a case, the constraints will be satisfied
at the boundary by virtue of the initial values alone, and there
is no need to impose additional conditions on the boundary values
in order to enforce them. The boundary values of the incoming
fields in this case must be arbitrary. Such formulations exist at
least in symmetry-reduced cases, and some boundary problems for
those formulations have been studied \cite{lecoiriondo}. 

Here we stray away from the general trend of seeking a way to
impose the constraints along the boundary, in order to propose a
seemingly unrelated method to write down equations that must hold
among the boundary values of many quite generic first-order
formulations of the Einstein equations. These consist of the
vanishing of the four components of the projection of the Einstein
tensor along the normal to the boundary.

In Section~\ref{sec:2} we briefly describe the formulation of the Einstein
equations that we choose to write down the proposed boundary
conditions explicitly in a model case. The discussion in Section~\ref{sec:2}
hinges heavily on the existence of a complete set of characteristic
fields, which is guaranteed by the strong hyperbolicity of the
formulation chosen (as opposed to weak hyperbolicity). In Section~\ref{sec:3}
we show that such boundary conditions are consistent with constraint
propagation, realize the goal of guaranteeing the vanishing of the
incoming constraints at the boundary and, furthermore, they coincide
with the boundary conditions proposed in \cite{calabrese01} by the
trading of space and time derivatives alluded to above. The
discussion on Section~\ref{sec:3} hinges heavily on the existence of a
complete set of characteristic constraint fields, namely: on the fact
that the constraint propagation is strongly hyperbolic (as opposed to
weakly hyperbolic). In Section~\ref{sec:4} we state the generality of the
arguments with regards to three-dimensional strongly hyperbolic
formulations, namely: the aspects of the argument that apply to any 
strongly hyperbolic formulation of the Einstein equations irrespective
of particularities such as the specific characteristic fields and
speeds.  Formulation-dependent features such as any explicit form of
the boundary conditions arising from the projection of the Einstein
equations perpendicularly to a boundary are intentionally excluded
from Section~\ref{sec:4} with the purpose of exposing the true reach and
relevance of the argument across the lines delimiting alternative
strongly hyperbolic formulations.  Concluding remarks appear in
Section~\ref{sec:5}.


\section{Boundary conditions for the Einstein-Christoffel
formulation with spherical symmetry \label{sec:2}}

As usual, in a foliation of spacetime by level surfaces of a time
function $t$, and using $t$ as a coordinate, the spacetime metric
can be viewed as a time dependent Riemannian metric evolving in
time according to a freely specifiable rate and labeling of the
3-space.  From now on, we use the term {\it metric} with no
qualifier to refer to the evolving Riemannian metric of the
slices. In this section we use the Einstein-Christoffel (EC)
formulation \cite{fixing}, restricted to spherical symmetry. The
full 3-D equations and their structure of characteristics can be
found in \cite{teukolsky}, as well as the restriction to spherical
symmetry. We borrow the notation directly from \cite{teukolsky}.
With the restriction to spherical symmetry, the spacetime metric
has the form
\begin{equation}
    ds^2 = -N^2dt^2+g_{rr}(dr+\beta^r dt)^2
		+ g_Tr^2(d\theta^2+\sin^2\theta d\phi^2)
\end{equation}

\noindent where $g_{rr}, g_T, \beta^r$ and $N$ are functions only
of $r$ and $t$. The shift $\beta^r$ and the densitized lapse
\begin{equation}
\widetilde{\alpha} \equiv \frac{N}{g_T\sqrt{g_{rr}}}
\end{equation}

\noindent are considered arbitrarily given. The EC formulation of
the Einstein equations restricted to spherical symmetry consists
of six fundamental variables
$(g_{rr},g_T,K_{rr},K_T,f_{rrr},f_{rT})$ which evolve according
to six evolution equations and whose initial data satisfies four
constraints \cite{teukolsky}. The evolution equations are
\numparts
\begin{eqnarray}
       \partial_t g_{rr} 
     - \beta^r\partial_r g_{rr}
 &=&
     -2NK_{rr} +2g_{rr}\partial_r\beta^r,\label{dotgrr}	\\
       \partial_t g_T 
     - \beta^r\partial_r g_T
 &=&
     -2NK_T +2\frac{\beta^r}{r} g_T,	\label{dotgT}
\end{eqnarray}
\begin{eqnarray}
\fl       \partial_t K_{rr} 
     - \beta^r\partial_r K_{rr}
	+\frac{N}{g_{rr}}\partial_r f_{rrr}
 &=&
      N\left[ 2f^r{}_{rr}\left(f^r{}_{rr} +\frac{1}{r}
			       -\frac{4f_{rT}}{g_T}\right)
	     +K_{rr}\left(2\frac{K_T}{g_T}-K^r_r\right)\right.\nonumber\\ &&\left.
	     -\frac{6}{r^2}      
     -6\left(\frac{f_{rT}}{g_T}\right)^2
     -\partial_r^2\ln \widetilde{\alpha}
     -(\partial_r\ln\widetilde{\alpha})^2 \right. \nonumber\\ 
\fl&& \left.
     +\left(\frac{4}{r}-f^r{}_{rr}\right)
	\partial_r\ln\widetilde{\alpha}
      \right]					
     +2K_{rr}\partial_r\beta_r \nonumber\\ \fl&&
     +4\pi N(Tg_{rr}-2S_{rr}),				\\
\fl       \partial_t K_T 
     - \beta^r\partial_r K_T
	+\frac{N}{g_{rr}}\partial_r f_{rT}
 &=&
      N\left( K_TK^r_r+\frac{1}{r^2}
	     -\frac{2f_{rT}^2}{g_{rr}{g_T}}
	     -\frac{f_{rT}}{g_{rr}}
	      \partial_r\ln\widetilde{\alpha}
       \right)	\nonumber\\ \fl&&
     +\frac{2\beta^r}{r}K_T,			\label{dotKT}\\
\fl       \partial_t f_{rrr} 
     - \beta^r\partial_r f_{rrr}
	+\frac{N}{g_{rr}}\partial_r K_{rr}
 &=&	
      N\left[ 4g_{rr}\frac{K_T}{g_T}
	      \left(3\frac{f_{rT}}{g_T}
		    -f^r{}_{rr} + \frac{2}{r}
		    -\partial_r\ln\widetilde{\alpha}
	      \right)\right.				\nonumber\\
\fl & &
	\left.
	     -K_{rr}\left(10\frac{f_{rT}}{g_T}
			  +f^r{}_{rr} - \frac{2}{r}
			  +\partial_r\ln\widetilde{\alpha}
		    \right)
        \right]					\nonumber\\
\fl & &  +3f_{rrr}\partial_r\beta^r+g_{rr}\partial_r^2\beta^r
     +16\pi N J_rg_{rr},
\\
\fl 	\partial_t f_{rT} 
     - \beta^r\partial_r f_{rT}
	+\frac{N}{g_{rr}}\partial_r K_T
 &=&	
      NK_T\left(2\frac{f_{rT}}{g_T}
		      -f^r{}_{rr}
		      -\partial_r\ln\widetilde{\alpha}		      
	      \right)\nonumber\\ \fl &&
     +\left(\partial_r\beta^r+\frac{2\beta^r}{r}\right)
       f_{rT}.					\label{dotfrT}
\end{eqnarray}
\endnumparts

\noindent The constraints are 
\numparts
\begin{eqnarray}
\fl	{\cal C}
&\equiv & 
	\frac{\partial_rf_{rT}}{g_{rr}g_T}
	-\frac{1}{2r^2g_t} 
        +\frac{f_{rT}}{g_{rr}g_T}\left(
	 \frac{2}{r} +\frac{7f_{rT}}{2g_T}-f^r{}_{rr}
			         \right)	 
-\frac{K_T}{g_T}\left(K^r_r+\frac{K_T}{2g_T}\right)
    +4\pi \rho =0, 
\end{eqnarray}
\begin{eqnarray}
\fl	{\cal C}_r
&\equiv &
	\frac{\partial_rK_T}{g_T}+\frac{2K_T}{rg_T}
        +\frac{f_{rT}}{g_T}\left(K^r_r+\frac{K_T}{g_T}\right)
    +4\pi J_r 			 = 0,	
\\				
\fl	{\cal C}_{rrr}
&\equiv &
	\partial_rg_{rr}
    +\frac{8g_{rr}f_{rT}}{g_T}-2f^r{}_{rr} =0,	\label{crrr}
\\
\fl	{\cal C}_{rT}
&\equiv &
	\partial_rg_T
    +\frac{2g_T}{r}-2f_{rT} =0.			\label{crT}
\end{eqnarray}
\endnumparts

\noindent Here ${\cal C}$ and ${\cal C}_r$ are the scalar and
vector constraints, respectively, whereas the vanishing of ${\cal
C}_{rrr}$ and ${\cal C}_{rT}$ defines the variables $f_{rrr}$ and
$f_{rT}$.  We are here interested only in the vacuum equations,
so we set all the source terms to zero ($T=S_{rr}=J_r=\rho=0$),
and we do not need to consider matter equations.    

In essence, the evolution system is strongly hyperbolic because one
can find a complete set of linearly independent combinations of the
fundamental variables such that the principal part of the equations
decouples.  Such combinations are referred to as the characteristic
variables. In a linear homogeneous system, the characteristic
variables propagate their initial values exactly along the
characteristic lines, and may be interpreted as waves propagating with
the characteristic speed. In our case \cite{teukolsky}, the
characteristic fields and their characteristic speeds are 
\numparts
\begin{eqnarray}
	U^0_r&\equiv& g_{rr} \hspace{0.5cm} (v_c = \beta^r)	\\
	U^0_T&\equiv& g_T    \hspace{0.5cm} (v_c = \beta^r)	\\
	U^\pm_r&\equiv& K_{rr}
		      \pm \frac{f_{rrr}}{\sqrt{g_{rr}}}
	\hspace{0.5cm}     (v_c = \beta^r\mp\widetilde{\alpha}g_T)\\
	U^\pm_T&\equiv& K_T
		      \pm \frac{f_{rT}}{\sqrt{g_{rr}}}
	\hspace{0.5cm}     (v_c = \beta^r\mp\widetilde{\alpha}g_T)\label{UTpm}
\end{eqnarray}	
\endnumparts

\noindent Here we have changed the signs of the characteristic speeds
relative to \cite{teukolsky} in order to agree with the sign
conventions in \cite{kreissbook}. The characteristic fields give
insights into how much data one can or must provide in addition to
the initial data, depending on the region of spacetime where the
solution is sought. If the region of interest has a boundary, in
general one may need to prescribe additional data on this boundary in
order to obtain a unique solution within that region. The additional
data are the values of the characteristic fields that enter the
region from outside. If such values are not prescribed, then there
are many solutions for the same initial data in the region of
interest. For each boundary value prescribed arbitrarily -- as long
as it is consistent with the initial data at the intersection of the
boundary with the initial slice --, there is a unique solution. No
boundary value may be freely prescribed for the characteristic fields
that cross the boundary from the inside, or that run along the
boundary, since such values are determined by the initial data
already, in principle. In practice, since the equations are not
linear nor homogeneous, the actual values of the outgoing
characteristic fields at the boundary are not known in advance. Most
of the time the values of some or even all the fields at the
boundary, regardless of formulation type, are prescribed by means of
some type of wave condition, as in \cite{shibnaka} and, more
elaborately, in \cite{admlong2}. 

In our problem, we restrict attention now to a region inside a fixed
value of $r$. There is a set of incoming characteristic fields, which
are the ones that have positive characteristic speeds in our
convention, which we will be able to single out once we choose the
values of the densitized lapse and shift. Suppose initial data are
prescribed with vanishing values of the constraints. One can now
prescribe the values of the incoming fields independently of the
initial values. The question is: are such values free? It is worth
investigating the possibility that the Einstein equations themselves
may restrict the boundary values. 

Consider the following argument. The boundary is a surface of
constant radius in spacetime. The unit normal vector to the
boundary is well defined. We denote it by
$e^a=(e^t,e^r,e^\theta,e^\phi)$ and it can easily be computed
from the gradient vector $r,_a = (0,1,0,0)$ by $e^a =
^4\!\!\!g^{ab}r,_b /\sqrt{^4\!g^{ab}r,_br,_a} =
^4\!\!g^{ar}/\sqrt{^4\!g^{rr}}$, where $^4\!g^{ab}$ is the
contravariant spacetime metric. In our case, we have 
\numparts
\begin{eqnarray}
	e^t 
&=& \frac{\beta^r}
       {\widetilde{\alpha}g_T
	\sqrt{g_{rr}(\widetilde{\alpha}^2g_T^2-(\beta^r)^2)}},\\
	e^r 
&=& \frac{\sqrt{\widetilde{\alpha}^2g_T^2-(\beta^r)^2}}
       {\widetilde{\alpha}g_T\sqrt{g_{rr}}},\\
	e^\theta &=& e^\phi = 0.
\end{eqnarray}
\endnumparts

\noindent The projector $p_{ab}$ on the boundary surface can be
found from the spacetime metric and the unit normal to the
surface via
\begin{equation}
	p_{ab} \equiv  \;{}^4\!g_{ab} - e_ae_b
\end{equation}

\noindent with latin indices $a,b,c\ldots$ raised and lowered
with the spacetime metric. In particular, we are interested in
the projector with mixed indices
\begin{equation}
	p^a{}_b = \delta^a{}_b - e^ae_b
\end{equation}

We saturate the indices of the Einstein equations $G_{ab}=0$ in
two different ways. One equation is obtained by contracting twice
with the normal to the boundary surface, $e^a$. We have
\begin{equation}\label{eGe}
G_{ab}e^ae^b = G_{tt}(e^t)^2+2G_{tr}e^re^t+G_{rr}(e^r)^2=0
\end{equation}

Another set of equations is obtained by contracting one idex with
$e^a$ and the other index with $p^b{}_c$, namely:
$G_{ab}e^ap^b{}_c=0$. But because of the spherical symmetry, this
contraction is identically zero for $c=\theta,\phi$.  The
remaining equations are 
\begin{eqnarray}
G_{ab}e^ap^b{}_t &=& G_{tt}e^t+G_{tr}e^r=0,	\label{eGpt}\\
G_{ab}e^ap^b{}_r &=& (G_{tt}e^t+G_{tr}e^r)p^t{}_r
	    +(G_{tr}e^t+G_{rr}e^r)p^r{}_r
=0. \label{eGpr}
\end{eqnarray}

\noindent Clearly, if (\ref{eGpt}) and (\ref{eGe}) are satisfied, 
then (\ref{eGpr}) is an identity.  Thus there are only two
independent equations, and we pick (\ref{eGpt}) and (\ref{eGe}). 
These equations are extremely valuable for the boundary value problem
because they contain no second-order derivatives with respect to $r$
irrespectively of the choice of $\beta^r$ or $\widetilde{\alpha}$. In
the case of a first-order formulation, like ours, such equations are
said to be {\it interior to the boundary surface} -- not to confuse
with the {\it interior} of our {\it region!} In the following, we
restrict ourselves to $\beta^r=0$ and $\widetilde{\alpha}=1$ just for
the sake of argument. Non trivial choices of densitized lapse and
shift only make the argument more complicated without introducing any
real obstacle. 

Explicitly, for the case of $\beta^r=0$ and
$\widetilde{\alpha}=1$, we have 
\numparts
\begin{eqnarray}
\fl
   G_{ab}e^ae_b = 0
&=& -\frac{1}{r^2g_T{}^2g_{rr}{}^2}
     \Big( 2r^2g_{rr}{}^{3/2}\dot{K}_T
           -3r^2g_{rr}f_{rT}{}^2  		
           -4rg_Tg_{rr}f_{rT}\nonumber\\
\fl &&
	   +2r^2g_Tf_{rT}f_{rrr}
	   -g_{rr}{}^2g_T		
	   +r^2g_{rr}K_T{}^2
      \Big), 		\label{eG'sa}\\
\fl    G_{ab}e^ap^b{}_t = 0
&=& \frac{2}{rg_Tg_{rr}{}^{3/2}}
    \Big( rg_{rr}\dot{f}_{rT} 
          +rg_T\sqrt{g_{rr}}f_{rT}K_{rr}	
          -\sqrt{g_{rr}}K_T
          \left( rg_{rr}f_{rT}\right.\nonumber\\ 
 \fl && \left.
		+2g_Tg_{rr} 
		-rg_Tf_{rrr}\right)\Big).\label{eG'sb}					\label{eGptfinal}
\end{eqnarray}
\endnumparts

\noindent We can put these equations entirely in terms of the
characteristic fields, using
\numparts
\begin{eqnarray}
   K_T     &=& \frac12\left(U^+_T+U^-_T\right),\label{kTcharac}\\
   f_{rT}  &=& \frac{\sqrt{g_{rr}}}{2}\left(U^+_T-U^-_T\right),
						\label{frTcharac}\\
   K_{rr}  &=& \frac12\left(U^+_r+U^-_r\right),	\\
   f_{rrr} &=& \frac{\sqrt{g_{rr}}}{2}\left(U^+_r-U^-_r\right).
\label{frrrcharac}
\end{eqnarray}
\endnumparts
			
\noindent and the fact that the metric components $g_{rr}$ and
$g_T$ are characteristic fields themselves. For the derivatives
$\dot{K}_T$ and $\dot{f}_{rT}$ we take derivatives of
(\ref{kTcharac}) and (\ref{frTcharac}) and use (\ref{dotgrr}) --
written in terms of the characteristic fields-- to substitute
$\dot{g}_{rr}$ which will appear in $\dot{f}_{rT}$. Explicitly,
for $\beta^r=0$ and $\widetilde{\alpha}=1$ we have
\numparts
\begin{eqnarray}
   \dot{K}_T 
&=& \frac12\left( \dot{U}^+_T +\dot{U}^-_T\right)
		\label{indotcharacteristicfieldsa}\\
   \dot{f}_{rT} 
&=& \frac{\sqrt{g_{rr}}}{2}
    \left(\dot{U}^+_T-\dot{U}^-_T\right)	
    -\frac{g_T}{4}\left(U^+_r+U^-_r\right)
		 \left(U^+_T-U^-_T\right)
\label{indotcharacteristicfieldsb}
\end{eqnarray}	
\endnumparts

\noindent On account of 
(\ref{indotcharacteristicfieldsa}-\ref{indotcharacteristicfieldsb}), 
one can see by inspection that
when both equations (\ref{eG'sa}-\ref{eG'sb}) are written out entirely in
terms of characteristic fields, they involve the time derivatives
only of $U^\pm_T$, but not of $U^\pm_r$. Furthermore,
$\dot{U}^\pm_T$ appear linearly in both equations. We can
algebraically solve the two equations for both $\dot{U}^\pm_T$ as
independent variables in terms of the rest.  We readily find:
\begin{eqnarray}\label{dotUT-}
	\dot{U}^-_T \!\!
  &=&\!\! \frac{1}{2r^2\sqrt{g_{rr}}}
    \Big(g_{rr}g_T
	 +r^2g_{rr}\left(U^-_T\right)^2
	-2r^2g_{rr}U^-_TU^+_T			\nonumber\\
&&	\!\!-4r\sqrt{g_{rr}}g_TU^-_T\!
	+r^2g_TU^-_TU^+_r\!
	-r^2g_TU^-_TU^-_r\Big)
\end{eqnarray}

\noindent and 
\begin{eqnarray}\label{dotUT+}
	\dot{U}^+_T \!\!
  &=&\!\! \frac{1}{2r^2\sqrt{g_{rr}}}
    \Big(g_{rr}g_T
	+r^2g_{rr}\left(U^+_T\right)^2
	-2r^2g_{rr}U^+_TU^-_T			\nonumber\\
&&	\!\!+4r\sqrt{g_{rr}}g_TU^+_T\!
	-r^2g_TU^+_TU^+_r\!
	+r^2g_TU^+_TU^-_r\Big)
\end{eqnarray}

\noindent Both equations (\ref{dotUT-}) and (\ref{dotUT+}) are
necessary because they are linearly independent. The question is:
how are we to interpret them within the framework of the
initial-boundary value problem? 

To start with, the fields $g_{rr}$ and $g_T$ may be considered as
known sources, because they travel upwards along the boundary. If
the problem was linear and homogeneous, then the values of  $g_{rr}$
and $g_T$ at any point on the boundary would be exactly their
initial values. Because the problem is non-linear, the values at any
point on the boundary must be integrated, but can still be regarded
as sources.  A similar thinking may be used for the outgoing
characteristic field $U^+_r$, with the complication that $U^+_r$
travels towards the boundary but not directly upwards, and thus the
computation of its values at the boundary will be more involved,
but conceptually no different.  

Additionally, we may take the point of view that the incoming
characteristic field $U^-_r$ is arbitrary, since there are only
two equations to satisfy at the boundary and $\dot{U}^-_r$ does
not appear in any of them.  From this perspective, the value of
this field at the boundary is truly free, and must be considered
as one true degree of freedom of the boundary value problem for
the Einstein equations, in addition to the degrees of freedom
contained in the initial data.  

Continuing within such an interpretation, Eq.~(\ref{dotUT-})
provides exact boundary values for the incoming characteristic field
$U^-_T$ in terms of the free incoming field $U^-_r$ and the fields
that either run along the boundary or cross from the inside, all of
which are determined by the initial data in principle, as argued
before.   

There remains Eq.~(\ref{dotUT+}). From our point of view, this
equation predicts boundary values for $U^+_T$. There is no doubt
that it must hold, in order for the solution that we are seeking
to satisfy all the Einstein equations at all points on the
boundary surface. But $U^+_T$ is, in principle, determined by the
initial data by propagation along characteristics, as explained
before. There seem to be two possibilities here: either the
outgoing field $U^+_T$ propagated from the initial data satisfies
Eq.~(\ref{dotUT+}), or it does not. If the boundary values of
$U^+_T$ do not satisfy Eq.~(\ref{dotUT+}), then the
initial-boundary value problem is inconsistent and {\it can
not\/} yield a solution to the Einstein equations in the region
of interest.

Alternatively, if the boundary values of $U^+_T$ propagated along
characteristics do satisfy Eq.~(\ref{dotUT+}), then the
initial-boundary value problem is consistent. In this case,
Eq.~(\ref{dotUT+}) can be used to prescribe the values of $U^+_T$
with more accuracy and reliability than the propagation along
characteristics discussed above because it is an ordinary
differential equation, and should be preferred. 

In the following Section we show that the initial-boundary value
problem is consistent and we should be using  Eq.~(\ref{dotUT+})
to prescribe the boundary values of $U^+_T$ instead of
propagating them by characteristics from the initial data,
because the propagation by characteristics is inaccurate in
itself --even more so in the case that the characteristic speeds
depend on the fields themselves and must be calculated at every
step. One can anticipate that this is the case based on some
intuition. Why should the boundary values of the outgoing fields
be ``constrained'' by an equation such as Eq.~(\ref{dotUT+})?
Because, in principle, the boundary values of the outgoing fields
must reflect in some way the constraints on the intial data that
determine them. The initial data are related; their
inter-relationships must be propagated along characteristics.
Eq.~(\ref{dotUT+}) may well be a shortcut out of constraint
propagation.

\section{Consistency with constraint propagation \label{sec:3}}

If thought of as functions
which could take any real values, the initial constraints ${\cal
C},{\cal C}_r,{\cal C}_{rrr}, {\cal C}_{rT}$ evolve in time
according to another strongly hyperbolic evolution system, with
the following characteristic fields and speeds\cite{calabrese01}:
\numparts
\begin{eqnarray}
	C_1 &=& {\cal C}+\frac{{\cal C}_r}{\sqrt{g_{rr}}}
	\hspace{0.5cm}(v_1^c = \beta^r-\widetilde{\alpha}g_T)	\\
	C_2 &=& {\cal C}-\frac{{\cal C}_r}{\sqrt{g_{rr}}}
	\hspace{0.5cm}(v_2^c = \beta^r+\widetilde{\alpha}g_T)	\\
	C_3 &=& {\cal C}_{rrr}
	\hspace{0.5cm}(v_3^c = \beta^r)	\\	
	C_4 &=& {\cal C}_{rT}
	\hspace{0.5cm}(v_4^c = \beta^r)	
\end{eqnarray}
\endnumparts

\noindent In the case of $\beta^r=0$ and $\widetilde{\alpha}=1$,
$C_2$ is the only incoming characteristic field. If one wishes for
all four functions to take the value zero in the region interior to
some fixed radius, then three of them will be zero by choice of
initial data, but $C_2$ must be set to zero at the boundary in order
to propagate inwardly from there. But one simply may set the value
of $C_2$ to zero because one does not evolve the constraint
propagation system.  It is one thing to impose the value $0$ on the
function $C_2$, but a different thing to expect $C_2$ to be
vanishing at the boundary as a function of the fundamental variables
of evolution. In particular, by inspection one can readily see that
$C_2$ contains the derivative of $U^-_T$ with respect to $r$.
Therefore, although it is natural and desirable to have $C_2=0$ on
the boundary, it is an impossible task to have $C_2=0$ as a boundary
condition for the fundamental fields of evolution. $C_2$ will have
to vanish for the evolution to produce a solution of the Einstein
equations, and thus the vanishing of $C_2$ and all the other
constraints must be checked for consistency at the boundary, after
the solution has been found and the $r-$derivatives can be
evaluated; but it is not practical as a boundary condition.    

A strategy that has been proposed \cite{calabrese01} to circumvent
this obstacle is to use the evolution equations to turn the
$r-$derivatives of the characteristic fields that appear in the
expression of $C_2$ into $t-$derivatives. The procedure must work in
this symmetry-reduced case because, by construction, all
characteristic fields have time derivatives proportional to their
$r-$derivatives, up to terms of zeroth order.  Therefore, $C_2=0$
can be turned into an evolution equation for $U^-_T$ that will be
restricted to the boundary. Following \cite{calabrese01}, we do that
next, maintaining the restrictions $\beta^r=0$ and
$\widetilde{\alpha}=1$. We have explicitly  
\begin{eqnarray}
\fl C_2&=& -\frac{1}{2r^2g_T{}^2g_{rr}{}^{5/2}}
\bigg(
	 2r^2g_Tg_{rr}\partial_rK_T
	-2r^2g_Tg_{rr}{}^{3/2}\partial_rf_{rT}
	+g_Tg_{rr}{}^{5/2}			\nonumber\\
\fl &&	-4rg_Tg_{rr}{}^{3/2}f_{rT}
	-7r^2g_{rr}{}^{3/2}f_{rT}		
	+2r^2g_T\sqrt{g_{rr}}f_{rT}f_{rrr}
	+2r^2g_Tg_{rr}{}^{3/2}K_TK_{rr}		\nonumber\\
\fl &&	+r^2g_{rr}{}^{5/2}K_T{}^2
	+4rg_tg_{rr}^2K_T
	-2r^2g_Tg_{rr}f_{rT}K_{rr}
	-2r^2g_{rr}^2f_{rT}K_T
\bigg)	\label{C2full}
\end{eqnarray}

\noindent Solving for $\partial_r f_{rT}$ from the evolution
equation (\ref{dotKT}), and for $\partial_r K_T$ from the
evolution equation (\ref{dotfrT}), substituting into
(\ref{C2full}), and replacing all appearances of
$f_{rT},f_{rrr},K_T$ and $K_{rr}$ in terms of the characteristic
fields using (\ref{kTcharac}-\ref{frrrcharac}), Eq.~(\ref{C2full})
turns into a new expression, which is not an initial constraint
anymore, so we use quotes to make this point explicit:
\begin{eqnarray}
\fl ``C_2{}" &=& -\frac{1}{2r^2g_T{}^2g_{rr}{}^{5/2}}
\left(
	 2r^2g_{rr}{}^2\dot{U}^-_T
	-r^2g_{rr}{}^{5/2}(U^-_T)^2
	-r^2g_Tg_{rr}{}^{3/2}U^-_TU^+_T	\right.	\nonumber\\
\fl &&
\left.	+r^2g_Tg_{rr}{}^{3/2}U^-_TU^-_r
	-g_Tg_{rr}{}^{5/2}
	+2r^2g_{rr}{}^{5/2}U^+_TU^-_T
	+4rg_Tg_{rr}{}^2U^-_T
\right)
\end{eqnarray}

\noindent Setting $``C_2{}"=0$ yields a boundary condition for
$U^-_T$.  The point is that $C_2=0$ on the boundary is not the
same as $``C_2{}"=0$, but they are the same along the boundary if
evaluated on fundamental fields that satisfy the evolution
equations. In other words: they differ by a linear combination of
the evolution equations. But more interestingly,  $``C_2{}"=0$ is exactly the
same equation as (\ref{dotUT-}), as can be verified by
inspection.  Therefore the method of ``trading'' space
derivatives for time derivatives advocated in \cite{calabrese01}
is equivalent to solving one of the Einstein equations that are
{\it interior to the boundary,\/} there being a two-dimensional
set of such equations. This could be expected because the projections
of the Einstein equations along the direction normal to a surface of
fixed radius are the linear combinations that have no
second-order $r-$derivatives, and clearly the ``trading'' is
equivalent to linearly combining the Einstein equations. Now,
because Eq.~(\ref{dotUT+}) is equivalent to $C_2=0$ on the boundary, it
guarantees that $C_2$ will vanish in the interior by propagation
along characteristics.  So Eq.~(\ref{dotUT+}) is not only
consistent with the boundary value problem, but is necessary as
well. 

We can also use the ``trading'' of space derivatives for time
derivatives to write an equation $``C_1{}"=0$ from $C_1=0$ along the
boundary. This is exactly Eq.~(\ref{dotUT+}), the meaning of which
is explained in the previous section, and we conclude that
Eq.~(\ref{dotUT+}) and $C_1$ are equivalent along the boundary if
the evolution equations are satisfied. But how much of this argument
proves that the constraint has been ``propagated'' and where does
the propagation start? Clearly $C_1$ propagates out from the initial
data towards the boundary, therefore $C_1$ will vanish along the
boundary as a consequence of its vanishing initially. Since the
evolution will be satisfied by construction, then $``C_1{}"$ will
vanish as a consequence of $C_1$ vanishing initially as well.
Seemingly, thus, the boundary values of the fields satisfying
Eq.~(\ref{dotUT+}) must be consistent with the initial values
satisfying $C_1=0$.  For this reason, and because Eq.~(\ref{dotUT+})
is an ordinary differential equation, it is not only consistent but
advisable to use Eq.~(\ref{dotUT+}) to obtain the values of $U^+_T$
instead of propagating them from the initial values along
characteristics. At any rate, Eq.~(\ref{dotUT+}) should be checked
for consistency regardless of method.

Two objections may come to mind on a surface glance to
Eq.~(\ref{dotUT+}), which we want to anticipate.  It might
conceivably be argued that if the values of $U^+_T$ at the boundary
are obtained from the initial values by propagation along
characteristics, then Eq.~(\ref{dotUT+}) could also be thought of as
an equation for the incoming field $U^-_r$ with $U^+_T$ given,
instead of prescribing $U^+_T$ with a free $U^-_r$.  But this is a
flawed line of thought, for the boundary values of $U^+_T$
propagated along characteristics are consistent with
Eq.~(\ref{dotUT+}), and substituting $U^+_T$ into Eq.~(\ref{dotUT+})
must yield an identity, leaving no equation to solve for $U^-_r$,
which would thus remain arbitrary.    

Secondly, how is it that Eq.~(\ref{dotUT+}) involving a free
boundary field $U^-_r$ can be consistent with propagation of $U^+_T$
along characteristics, given that $U^-_r$ is a field which the
initial data know nothing about? The initial data do not know about
$U^-_r$, but the characteristics do, because the problem is non
linear. Thus the propagation of $U^+_T$ along characteristics will
necessarily yield boundary values of $U^+_T$ that know about earlier
boundary values of incoming fields. It is fine for
Eq.~(\ref{dotUT+}) to involve $U^-_r$, as long as it does not
involve $\dot{U}^-_r$.  

For additional support of our scheme to use Eq.~(\ref{dotUT-})
and  Eq.~(\ref{dotUT+}) to prescribe $U^-_T$ and $U^+_T$,
respectively, assuming that $U^-_r$ is given freely, we can
linearize the equations around flat space and verify that the
scheme is consistent, since the boundary value problem of linear
hyperbolic equations is clear and enlightening. We have
\numparts
\begin{eqnarray}
	g_{rr} &=&           1+\widehat{g}_{rr}	\\
	g_T    &=&           1+\widehat{g}_T	\\
	K_{rr} &=&             \widehat{K}_{rr}	\\
	K_T    &=&             \widehat{K}_T	\\
	f_{rrr}&=& \frac{4}{r}+\widehat{f}_{rrr}	\\
	f_{rT} &=& \frac{1}{r}+\widehat{f}_{rT}	
\end{eqnarray}
\endnumparts

\noindent where all hatted quantities are small.  We will keep
only linear terms in such quantities. The evolution equations
become
\numparts
\begin{eqnarray}
\dot{\widehat{g}}_{rr} &=& -2\widehat{K}_{rr}	\\
\dot{\widehat{g}}_T    &=& -2\widehat{K}_T	\\
\dot{\widehat{K}}_{rr} 
+\partial_r\widehat{f}_{rrr} 
&=&
-\frac{42}{r^2}\widehat{g}_{rr}
+\frac{48}{r^2}\widehat{g}_T
+\frac{10}{r}  \widehat{f}_{rrr}	
-\frac{44}{r}  \widehat{f}_{rT},			\\
\dot{\widehat{K}}_T 
+\partial_r\widehat{f}_{rT} 
&=&
 \frac{1}{r^2}\widehat{g}_{rr}
+\frac{2}{r^2}\widehat{g}_T
-\frac{4}{r} \widehat{f}_{rT}			\\
\dot{\widehat{f}}_{rrr} 
+\partial_r\widehat{K}_{rr} 
&=&
-\frac{12}{r}\widehat{K}_{rr}
+\frac{4}{r}\widehat{K}_T			\\
\dot{\widehat{f}}_{rT} 
+\partial_r\widehat{K}_T 
&=&
-\frac{2}{r}\widehat{K}_T			
\end{eqnarray}
\endnumparts

\noindent Thus the characteristic fields are $\widehat{g}_{rr}, 
\widehat{g}_T$ and 
\numparts
\begin{eqnarray}
\widehat{U}^\pm_T &\equiv& \widehat{K}_T\pm \widehat{f}_{rT},\\
\widehat{U}^\pm_r &\equiv& \widehat{K}_{rr}\pm \widehat{f}_{rrr}.
\end{eqnarray}
\endnumparts

\noindent Our boundary conditions, Eqs.~(\ref{dotUT-}) and 
(\ref{dotUT+}), linearize to
\begin{equation}
\dot{\widehat{U}}{}^-_T
= \frac{9}{2r^2}\widehat{g}_{rr}
   -\frac{3}{2r^2}\widehat{g}_T
   +\frac{1}{r}   \widehat{U}^-_r
   +\frac{1}{r}   \widehat{U}^+_T\; ,	\label{lindotUT-}
\end{equation}

\noindent and 
\begin{equation}
\dot{\widehat{U}}{}^+_T
= \frac{9}{2r^2}\widehat{g}_{rr}
   -\frac{3}{2r^2}\widehat{g}_T
   -\frac{1}{r}   \widehat{U}^+_r
   -\frac{1}{r}   \widehat{U}^-_T\; ,	\label{lindotUT+}
\end{equation}

\noindent respectively. Thus (\ref{lindotUT-}) can be used to
prescribe values for $\widehat{U}^-_T$ if $\widehat{U}^-_r$ is
given arbitrarily, whereas (\ref{lindotUT+}) can be used to
calculate values for  $\widehat{U}^+_T$ irrespective of
$\widehat{U}^-_r$, as anticipated.

\section{Boundary conditions for generic three-dimensional
strongly hyperbolic formulations \label{sec:4}}

How much of the argument in spherical symmetry actually depends
on the symmetry assumption? Not a great deal.  Suppose we have a
strongly hyperbolic formulation of the Einstein equations in
terms of 6+6+18 variables representing the three-metric and all
its first derivatives. The evolution equations then look like
\begin{equation}\label{evolgen}
	\dot{u} = A^i\partial_i u + b
\end{equation}

\noindent where $u$ is the 30-dimensional vector of all the
fundamental variables, and there are 4+18 constraints on the
initial data:
\numparts\label{constgen}
\begin{eqnarray}
	{\cal C}       &=& 0	\\
	{\cal C}_i     &=& 0	\\
	{\cal C}_{ijk} &=& 0	
\end{eqnarray}
\endnumparts

\noindent where ${\cal C}$ and ${\cal C}_i$ are the scalar and
vector constraint, respectively, and ${\cal C}_{ijk}$ are the
constraints necessary to reduce the equations from second to
first order in space (they define the 18 first order variables).
Wherever the set of evolution equations and the constraints are
satisfied, the ten Einstein equations $G_{ab}=0$ for the ten
components of the spacetime metric $g_{ab}$ are satisfied,
equivalently. 

Suppose the evolution equations (\ref{evolgen}) are strongly
hyperbolic, and that they imply, in the usual
manner~\cite{simoconst}, a second (or {\it subsidiary\/}) system of
equations for the constraint functions ${\cal C},{\cal C}_i$ and
${\cal C}_{ijk}$. Suppose that this second system of equations, for
the constraints, is also strongly hyperbolic. Assume we have
identified the characteristic variables of both strongly hyperbolic
systems. 

Suppose now that we seek a solution in the region interior to
some fixed value of the coordinate $x^1$ (any spacelike
coordinate). The unit normal to the boundary is then
\begin{equation}
	e^a = \frac{g^{ab}\delta^1_b}
		   {\sqrt{g^{ab}\delta^1_a\delta^1_b}}
	    = \frac{g^{1a}}{\sqrt{g^{11}}}
\end{equation}

\noindent and the projector on the boundary surface is $p^a{}_b =
\delta^a_b+e^ae_b$ with $e_b = \delta^1_b/\sqrt{g^{11}}$. We can
write down $G_{ab}e^ae^b = 0$ and $G_abe^ap^b{}_c = 0$.  In
general these will be four equations with {\it no second
derivatives with respect to} $x^1$, as can be proven by direct
calculation.  Contracting the Einstein tensor with $e^a$ yields a
vector equation with two pieces:
\begin{equation}
G_{ab}e^a  = R_{ab}e^a - \frac{\delta^1_b}{2\sqrt{g^{11}}}R .
\end{equation}

\noindent We next show that the components $b\neq 1$ of
$R_{ab}e^a$ have no second derivatives of the metric with respect
to $x^1$, and that the Ricci scalar term cancels out the second
$x^1-$derivatives that appear in the first term for $b=1$.  We
use the following well-known expression of the Ricci tensor
\cite{weinberg} in which the second-order derivatives of the metric
are explicit:
\begin{equation}
\fl	R_{ab} 
   = \frac12 g^{cd}
	\left( g_{cb,ad} +g_{ad,cb} -g_{cd,ab} -g_{ab,cd}\right)						
	+g^{cd}\left( \Gamma^e_{ad}\Gamma_{edb}
		     -\Gamma^e_{ab}\Gamma_{ecd}\right).
\end{equation}

\noindent Contracting with $e^a$ yields
\begin{equation}
R_{ab}e^a = 
\frac{1}{2\sqrt{g^{11}}}
 \left(g^{1a}g^{1c}g_{ac,1b} 
		-g^{11}g^{cd}g_{cd,1b}\right)
+\ldots
\end{equation}

\noindent where $\ldots$ represents terms that have no second
derivatives with respect to $x^1$.  Thus $R_{ab}e^a$ has no
second $x^1-$derivatives except for $b=1$.  The second part of
the Einstein tensor is nonvanishing only for $b=1$.  Therefore
the only component of $G_{ab}e^a$ that might contain a second
$x^1-$derivative is $b=1$.  Now $R$ is exactly 
\begin{equation}
	R = g^{1a}g^{1c}g_{ac,11}-g^{11}g^{cd}g_{cd,11} 
	    +\ldots
\end{equation}

\noindent Therefore, the terms with second $x^1-$derivatives of
the Ricci scalar part of the Einstein tensor cancel exactly with
those in the Ricci tensor part, with the consequence that
$G_{ab}e^a$ has no second  $x^1-$derivatives of the metric for
any value of $b$.  Contracting with $e^b$ or with $p^b{}_c$ will
not add new second  $x^1-$derivatives, but provides perhaps a
convenient splitting of the four equations. The four
equations
\begin{equation}\label{boundarygen}
	{\cal B}_b \equiv G_{ab}e^a =0 
\end{equation}
or equivalently
\begin{equation} 
	G_{ab}e^ae^b=0 \;\;\mbox{ and }\;\;
	G_{ab}e^ap^b{}_c=0
\end{equation}

\noindent must be satisfied by any solution to the Einstein
equations and are {\it interior\/} to the boundary surface, which
makes them ideal boundary conditions. There are four equations and
thirty variables, but of the thirty variables many are outgoing or
run along the boundary.  In fact, one can always write the evolution
system so that the metric components propagate with vanishing
characteristic speeds, regardless of the choice of lapse and shift.
So we have at most $(30-6)/2$ relevant incoming characteristic
fields. Clearly, some of the incoming fields will be left arbitrary
at the boundary -- perhaps more than 8, since in our example in
spherical symmetry we find that one of the boundary conditions
applies to an outgoing field instead of an incoming one. The
particulars of which characteristic fields will be affected by the
boundary equations will depend entirely on the details of the
first-order strongly hyperbolic formulation of choice. As an
illustration, the direct generalization of our argument to the case
of the Einstein-Christoffel formulation in three-dimensions without
the restriction of spherical symmetry will be reported elsewhere. 

Conceptually, the scheme runs as follows.  Choose initial data
that satisfy (\ref{constgen}). Choose boundary data that satisfy
(\ref{boundarygen}). The vanishing of the initial constraints
guarantees that the outgoing and static constraints will remain
vanishing in the region of interest. The boundary equations
(\ref{boundarygen}) are equivalent to the vanishing of the
incoming constraints at the boundary on account of their being
related to the incoming constraints by linear combinations with
the Einstein equations that {\it are\/} satisfied at the boundary
-- the evolution is satisfied by construction and the outgoing
constraints are satisfied by propagation along characteristics.
Therefore the boundary equations (\ref{boundarygen}) guarantee
that the incoming constraints are vanishing at the boundary,
which in turn guarantees that they will remain vanishing in the
region of interest.   

Although the preceding calculation assumes that the normal to the
surface of fixed value of $x^1$ is spacelike, clearly
Eqs.~(\ref{boundarygen}) will be interior to the surface of fixed
value of $x^1$ even if the normal is timelike, in which case the
proof that the equations contain no second $x^1-$derivatives runs
the same if we substitute $\sqrt{g^{11}}$ with  $\sqrt{|g^{11}|}$.
This remark applies to interior boundaries that lie within the event
horizon of a black hole spacetime, and may be relevant to numerical
simulations of binary black holes.  

On the other hand, in this Section we have restricted the discussion
to three-dimensional boundaries taken one at a time. As is the case
with any three-dimensional boundary-value problem, the intersection
of two boundaries requires special attention. In this respect,
consistency issues between the boundary equations that may arise at
the intersection of two boundary surfaces remain to be studied, and
their resolution may depend on the particulars of the formulation of
the initial value problem at hand. The interested reader is referred
to ~\cite{roberadm} where a particular representation of the
projection of the Einstein equations on the boundary is implemented
and corners and edges are treated, for the case of an ADM-like
formulation of the initial value problem.

\section{Concluding remarks and outlook \label{sec:5}}

Even though we have, for the most part, developed our arguments
explicitly in the case of spherical symmetry for the EC formulation,
the fundamental relevance of the argument to the initial-boundary
value problem of the Einstein equations does not depend on the
symmetry restrictions nor on the particulars of the hyperbolic
formulation, as we show in Section~\ref{sec:4}.  

We argue that given any boundary at a fixed value of a coordinate
of an initial-boundary value problem for the Einstein equations,
the vanishing of the components of the projection of the Einstein
tensor $G_{ab}$ along the normal $e^a$ to the boundary, namely
$G_{ab}e^b \equiv {\cal B}_a=0$, constitute necessary and
consistent boundary conditions, for essential reasons, as follows. 
  
First, the components of the projection of the Einstein tensor along
the normal to such a boundary, ${\cal B}_a$, contain no second-order
derivatives sticking out of the boundary.  If the initial value
problem is stated in first order form, such as any of the hyperbolic
formulations available, their vanishing becomes differential
equations for the boundary values of the fundamental fields
excluding their derivatives across the boundary. Moreover, if the
initial value problem is stated in second-order form in the space
coordinates, such as ADM form \cite{ADM,yorksources} or conformal
form \cite{shibnaka,BSSN}, they constitute a type of mixed
Neumann-Dirichlet conditions, the appropriateness of which is worth
investigating in full depth. In this regard, some representations of
${\cal B}_a=0$ as boundary conditions have been used along with an
ADM type formulation of the linearized Einstein equations in
\cite{roberadm} to investigate the consistency of pure Neumann or
Dirichlet conditions with the evolution.  

Second, in the case of strongly hyperbolic formulations that
propagate the constraints in a strongly hyperbolic fashion, the
vanishing of all four boundary equations ${\cal B}_a$ guarantees the
vanishing of the incoming constraints in the region of interest and
is consistent with the propagation of the outgoing constraints along
characteristics. In practice, this means that one can use ${\cal
B}_a=0$ to prescribe as many boundary values as possible, without
worrying about inconsistencies with the initial values.  This
necessary analytic consistency should not be mistaken for numerical
consistency, however. In fact, it is not known at this stage whether
this analytic consistency is stable under small perturbations of
either the initial data or the boundary data, a point that is
critical to the numerical implementation. Additionally, how to
proceed in order to investigate the stability of an
initial-boundary-value problem with constraints remains, to our
knowledge, an open problem not addressed in the standard reference
literature of strongly hyperbolic systems such as
\cite{kreissbook}.   

In the case of formulations that do not guarantee stable constraint
propagation, ${\cal B}_a=0$ on the boundary still appeals to us as
the next best thing. In fact, in such cases there are hardly any
handles on boundary conditions. Notice that imposing ${\cal B}_a=0$
on the boundary does not involve any manipulation of the choice of
evolution equations; it does not affect the evolution equations
themselves nor the system of propagation of the constraints. The
converse is not true: any manipulation of the evolution equations (by
adding linear combinations of the constraints in the usual manner)
necessarily affects both the propagation of the constraints and the
role of ${\cal B}_a=0$ on the boundary.  The interested reader is
referred to \cite{hisaaki} for a series of numerical studies of the
effect of the manipulation of the evolution equations on constraint
propagation exclusively, without imposing ${\cal B}_a=0$.   

Several issues remain open at this time. First and foremost, there
remains the issue of whether the boundary equations ${\cal B}_a=0$
are consistent with a well-posed initial-boundary-value problem in
the analytic sense, that is: in the sense that the solutions at a
later time are continuous functions of the initial data and the
boundary free data. The answer to this question will vary among the
different strongly hyperbolic formulations available. The answer may
be relevant to numerical relativity because, even though
well-posedness of the initial-boundary value problem does not
guarantee a stable numerical implementation, it is considered as a
necessary~\cite{kreissbook} or at least desirable feature of the
continuous equations being implemented. Second, any issues that may
arise exclusively in connection with the numerical implementation of
${\cal B}_a=0$ --but that are otherwise irrelevant to the analytic
initial-boundary value problem of the Einstein equations-- are also
open at this time and are worth pursuing, their details being strongly
dependent on the particulars of the formulation at hand. From the
numerical point of view, for instance, the presence of
undifferentiated terms, --which is not relevant to our current
argument and does not affect the well-posedness of first-order
problems-- is critical to numerical stability ~\cite{kreissbook}. In
this respect, the effect of the undifferentiated terms of the
boundary equations ${\cal B}_a=0$ on any issues of stability remains
to be determined, being, almost certainly, critically dependent on
the particulars of the formulation at hand. Additional open issues at
this time include checking for the consistency of ${\cal B}_a=0$ with
other boundary conditions already in use in numerical simulations,
and, eventually, how the use of ${\cal B}_a=0$ on the boundary may
affect the run time of numerical simulations.

\ack
This work was supported by the NSF under grants No. PHY-0070624
to Duquesne University and No. PHY-0135390 to Carnegie Mellon
University.


\end{document}